\documentclass[apj]{emulateapj}
\usepackage{apjfonts}

\begin{document}


\title{A Connection Between Bulge Properties and the Bimodality of
  Galaxies}

\shorttitle{Bulge properties and the bimodality of galaxies}
\shortauthors{Drory \& Fisher}


\author{Niv~Drory\altaffilmark{1}}
\author{David~B.~Fisher}

\affil{Department of Astronomy, The University of Texas at Austin,\\
  1 University Station C1400, Austin, Texas 78712}
\email{drory@astro.as.utexas.edu}

\altaffiltext{1}{Current address: Max-Planck-Institut f\"ur
  Extraterrestrische Physik, Giessenbachstra\ss e, Garching, Germany}

\slugcomment{Submitted to ApJ}


\begin{abstract}
  The global colors and structural properties of galaxies have
  recently been shown to follow bimodal distributions. Galaxies
  separate into a ``red sequence'', populated prototypically by
  early-type galaxies, and a ``blue cloud'', whose typical objects are
  late-type disk galaxies. Intermediate-type (Sa-Sbc) galaxies
  populate both regions. It has been suggested that this bimodality
  reflects the two-component nature of disk-bulge galaxies. However,
  it has now been established that there are two types of bulges:
  ``classical bulges'' that are dynamically hot systems resembling
  (little) ellipticals, and ``pseudobulges'', dynamically cold,
  flattened, disk-like structures that could not have formed via
  violent relaxation. Alas, given the different formation mechanisms
  of these bulges, the question is whether at types Sa-Sbc, where both
  bulge types are found, the red-blue dichotomy separates galaxies at
  some value of disk-to-bulge ratio, $B/T$, or, whether it separates
  galaxies of different bulge type, irrespective of their $B/T$.

  In this paper, we identify classical bulges and pseudobulges
  morphologically with HST images in a sample of nearby
  galaxies. Detailed surface photometry reveals that: (1) The red --
  blue dichotomy is a function of bulge type: at the same $B/T$,
  pseudobulges are in globally blue galaxies and classical bulges are
  in globally red galaxies. (2) Bulge type also predicts where the
  galaxy lies in other (bimodal) global structural parameters: global
  S\'ersic index and central surface brightness. (3) Hence, the red --
  blue dichotomy is not due to decreasing bulge prominence alone, and
  the bulge type of a galaxy carries significance for the galaxy's
  evolutionary history.

  We interpret this result as showing that the type of bulge a galaxy
  has is a signpost of the evolutionary history of the whole
  galaxy. Classical bulges are thought to indicate that a galaxy has
  undergone violent relaxation, e.g.\ during a major merger (of
  smaller fragments) in its past. This is more likely to have happened
  earlier when merging was frequent, in higher-density environments,
  and when there was still enough gas to subsequently form the disk.
  Therefore, these galaxies are likely to be red today.  Pseudobulges
  are disk components and therefore indicate a disk-only galaxy.  Such
  a galaxy has not suffered a major merger since the formation of its
  disk. This is more likely at later epochs, when the merger rate is
  lower and in low-density environments. Therefore, these are likely
  to be younger, blue galaxies. In addition, there is evidence that
  pseudobulge galaxies harbor supermassive black holes that follow the
  $M_{BH}$--$\sigma$ relation. We discuss the effect of black hole
  feedback in the host galaxy. If feedback during black hole growth in
  classical bulges is what quenches star formation in their host
  galaxies, why does this not happen in galaxies with pseudobulges?
\end{abstract}



\section{Introduction}\label{sec:introduction}

The study of the statistical properties of galaxies has advanced
rapidly due to the release of large amounts of homogeneously sampled
data from wide area surveys such as the Sloan Digital Sky Survey
(SDSS; \citealp{SDSS}).  The most notable result of recent years has
been the bimodal distribution of many galaxy properties. Galaxies are
bimodally distributed in the color--magnitude plane separating into a
red sequence and a blue cloud
\citep{Stratevaetal01,Baloghetal04,Baldryetal04} and also
\citep{Liskeetal03,Driveretal06}. Other structural parameters, such as
luminosity, size, surface density, and concentration, separate
similarly into two sequences
\citep{Blantonetal03,Shenetal03,Kauffmannetal03b,Driveretal06}.
Stellar populations and stellar masses and mass surface density follow
as well \citep{Kauffmannetal03a,Kauffmannetal03b}.

It is worth pointing out that some aspects of this bimodality in
galaxy properties have been known in other forms for a while. An
example is the existence of a tight color--magnitude relation for
early-type galaxies and the non-existence of such a relation for
late-type galaxies \citep{dVdV73}.

The fact that the bimodality is manifested in many parameters in a
similar way is perhaps not surprising. Structural and stellar
population related parameters of galaxies are known to be well
correlated, giving rise to galaxy classification schemes such as the
Hubble Sequence (\citealp{Hubble26,Sandage61}; see \citealp{RH94} for
a review of parameter correlations along the Hubble Sequence).

On the one extreme of the Hubble Sequence we find elliptical (E)
galaxies, which are thought to be the prototypical red-sequence
objects, and on the other extreme pure disk galaxies (Sd-Sm), which
populate the blue cloud. Intermediate-type (Sa-Sbc) galaxies form a
sequence in bulge-to-total ratio, $B/T$, and therefore bridge the red
and blue loci in the color--magnitude plane. It is therefore reasonable
to attribute the bimodality seen in colors of galaxies to this
bulge-disk two-component nature of galaxies, a point recently affirmed
by \citet{Driveretal06}.

Yet, identifying the physical structures that are responsible for the
bimodal distribution is not entirely straight-forward. Firstly, colors
of disks and their associated bulges are correlated, such that redder
disks harbor redder bulges
\citep{PB96,deJong96a,MacArthur04}. Secondly, it has now been
established that there are at least two types of bulges, where
``bulge'' is defined as the excess light over the inward extrapolation
of the surface brightness profile of the outer disk. The common
procedure in the literature to identify bulges is by surface
brightness profile decomposition and this practice identifies all
types of bulges in a common fashion. Thus we still refer to all the
structures that are found in this way as ``bulges''. A more physically
motivated definition is given below, however the term ``bulge''
defined in such purely observational terms is still operationally
useful and hence we will adopt this photometric definition in this
paper. We will, however, prefix this observationally-motivated term
``bulge'' by physically-motivated qualifiers.

Many bulges are dynamically hot systems resembling elliptical galaxies
that happen to have a disk around them \citep[e.g.][]{Renzini99}.
These bulges are called ``classical bulges''. Their formation is
assumed to be similar to that of elliptical galaxies, which are
thought to be the products of violent relaxation during major
mergers. This happens in environmentally driven evolution
(hierarchical clustering), which was the dominant process driving
galaxy formation in the early universe.

On the other hand, ``pseudobulges'' are bulges that have structure and
kinematics resembling that of disks. They are believed to have formed
via dramatically different formation channels than those responsible
for the formation of classical bulges (see \citealp{KK04} for a
comprehensive review). Pseudobulges are dynamically cold
\citep{Kormendy93}. They have flattening similar to that of the outer
disk \citep{Kent85,Kormendy93,FP03,KF05,KCBKA06}. Also, they may have
embedded secondary bars, rings, and/or spiral structure
\citep{carollo1997}. All these are disk phenomena which do not occur
in hot stellar systems. Therefore, these bulges could not have been
formed by mergers involving violent relaxation. Instead, they are
thought to form through slow rearrangement of disk material. Disk
galaxies form their structural analogs to these nuclear features as a
result of having high angular momentum compared to random motions.  We
expect a similar phenomenon is occurring in pseudobulges.
\citet{Kormendy93} shows that some bulges do have stellar dynamics
which resemble inclined disks better than they do oblate rotators.

What can drive disk galaxies to reshape themselves to form a
pseudobulge?  Non-axisymmetries in the gravitational potential (bars,
ovals, spiral structure) redistribute energy and angular momentum in
disks. A large body of literature reviewed by \citet{KK04} makes a
strong case that bars rearrange disk gas into inner rings, outer
rings, and dump gas into the center. All indications are that internal
evolution builds up the central gas density, resulting in star
formation and bulge-like stellar densities, thus forming
pseudobulges. Internal evolution is mostly ``secular'' -- its
timescales are much longer than the collapse time. \cite{Fisher06}
shows that pseudobulges are experiencing enhanced star formation over
classical bulges.  This further favors the idea that pseudobulges form
through a slower process (secular evolution), and hence are still
being built today. For recent reviews on this subject see \citet{KK04}
and \citet{Atha05}.

Hence, if pseudobulges are truly disk phenomena (in some way like bars
or rings), it is natural to expect that the dichotomy of galaxy
properties is not merely a product of changing bulge-to-total ratio,
but distinguishes {\em disks} (including their pseudobulges) from {\em
  classical bulges}. This imposes us to ask two questions. Do galaxies
with pseudobulges behave like bulgeless pure disk galaxies? Secondly,
is the location of a galaxy with respect to the (color) bimodality
determined by the relative prominence of its bulge and disk components
alone?

The existence of a dichotomy among bulges themselves offers an
opportunity to test this. The question becomes whether at intermediate
Hubble types of Sa-Sbc, where both bulge types are found, the color
bimodality separates galaxies at some bulge-to-total ratio, or, whether
it separates galaxies of different bulge type, irrespective of
bulge-to-total ratio (or neither).

In this paper, we answer these questions by comparing the location of
galaxies with pseudobulges to that of galaxies with classical bulges
with respect to the bimodality of the color and structural
distributions of galaxies in general, i.e.\ we look for a relationship
between the type of bulge a galaxy has and the global properties of
the galaxy harboring the bulge.

This paper is organized as follows. Sect.~\ref{sec:galaxy-sample}
discusses the galaxy sample used in this work and lays out the
analysis methods and bulge classification scheme we use. In
Sect.~\ref{sec:bulge-type} we present and discuss the dependence of
galaxy properties on bulge type. Finally, we summarize and discuss our
results in Sect.~\ref{sec:summary}.


\section{The galaxy sample}\label{sec:galaxy-sample}

The aim of this work is to compare the location of galaxies with
pseudobulges to that of galaxies with classical bulges with respect to
the bimodal distribution of global galaxy colors and structural
properties.  We will compare the loci of galaxies with pseudobulges to
that of galaxies with classical bulges in the color--magnitude plane
($u\!-\!r$ vs.\ $M_r$) and structure-color plane (central surface
brightness, $\mu_0$, and global S\'ersic index, $n$, vs.\ $u\!-\!r$).

We select a sample of 39 galaxies spanning Hubble types S0 to Sc by
cross referencing the Third Reference Catalog of Bright Galaxies (RC3;
\citealp{RC3}), the Sloan Digital Sky Survey Data Release Four (SDSS -
DR4) database \citep{SDSS-DR4}, and the Hubble Space Telescope (HST)
archive. We require that the galaxies have inclination $i \leq
60\degr$ to reduce the effect of dust. We will use the RC3 Hubble
classification, colors and total magnitudes from SDSS images, and
surface brightness profile fits to combined HST and SDSS surface
photometry. We identify pseudobulges and classical bulges using the
high-resolution HST images. We maintain a roughly even sampling of
Hubble types from S0 to Sc.

\subsection{Identification of pseudobulges}\label{sec:ident-pseudo}

\begin{figure*}[t]
  \centering
  \includegraphics[width=0.85\textwidth]{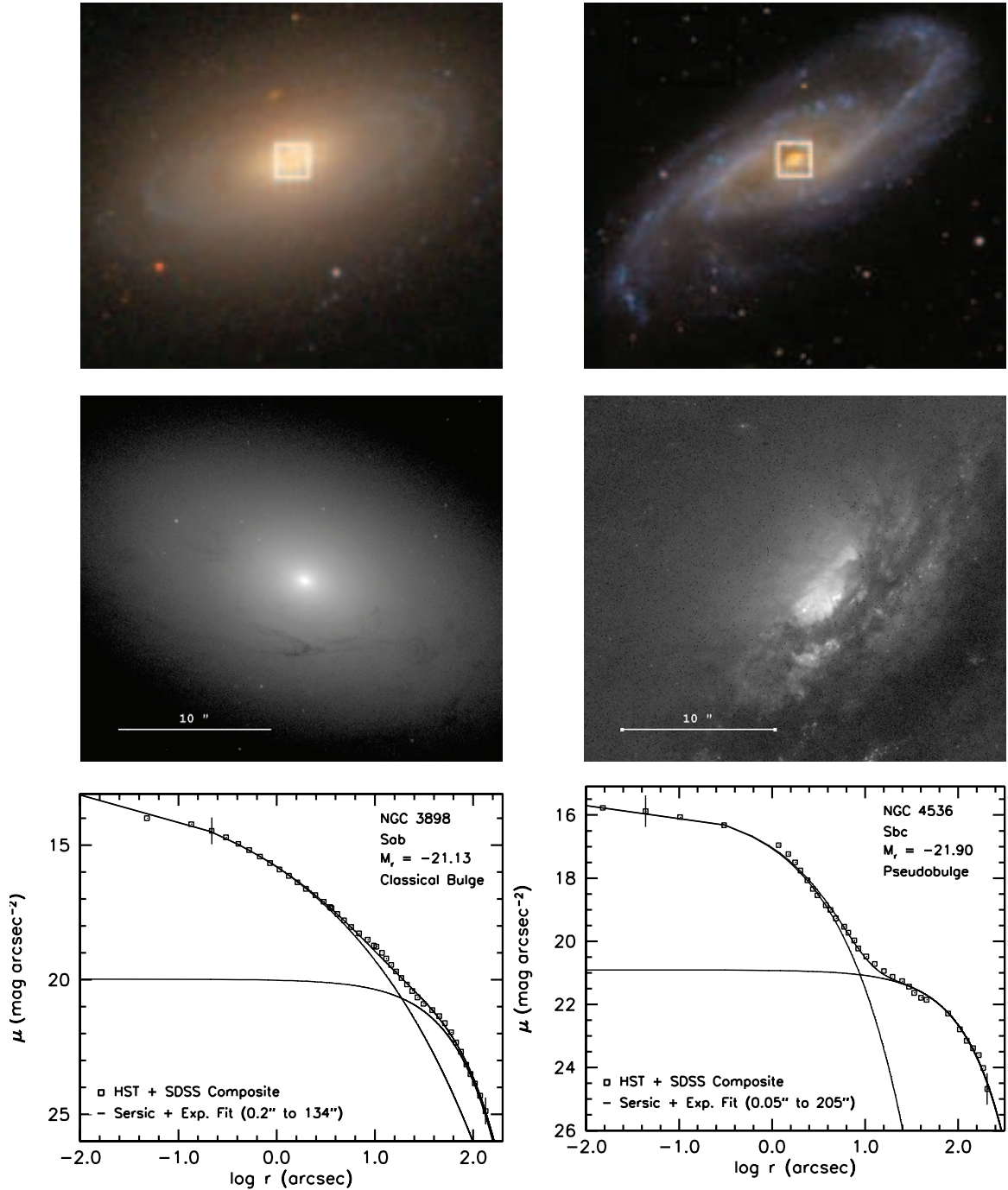}
  \caption{An example of pseudobulge (NGC~4536; right panels) and
    classical bulge (NGC~3898; left panels) identification. The top
    panels show an SDSS color image of each galaxy. The middle panels
    show HST WFPC2 images of the bulges of their respective
    galaxy. The white box on the SDSS image shows the extent of the
    HST images, and 10\arcsec\ is marked as a white line in the HST
    images for scale. Bulge-disk decompositions of composite HST plus
    SDSS surface brightness profiles are shown in the bottom
    panels.}\label{fig:bulges}
\end{figure*}

In this study, we classify galaxies as having a pseudobulge using
bulge morphology; if the ``bulge'' is or contains a nuclear bar,
nuclear spiral, and/or nuclear ring the ``bulge'' is actually a
pseudobulge. Conversely if the bulge is featureless and more round
than the outer disk, the bulge is called a classical
bulge. Fig.~\ref{fig:bulges} illustrates typical examples of what we
identify as classical bulges (left panels) and a pseudobulges (right
panels). Notice first, that the classical bulge (NGC~3898) has a
smooth and regular stellar light profile as seen in the WFPC2 F547M
image in the middle panel of Fig.~\ref{fig:bulges}. There is no reason
evident in the image to think that this galaxy harbors a
pseudobulge. On the other hand, NGC~4536 is a typical example of a
galaxy with nuclear spiral structure and patchiness (i.e. a
pseudobulge).  A decomposition of the stellar surface brightness
profile shows that the pseudobulge dominates the light profile to a
radius of $\sim 9$ arcseconds.  The WFPC2 F606W image in the middle
panel implies that the entire pseudobulge appears to exhibit spiral
structure. Notice that spiral structure exists in the small region
inside the box on the wider field SDSS image.  Also notice that the
classical bulge has a $\mu(r) \propto r^{1/3.4}$ profile, while the
pseudobulge is nearly exponential.

We identify pseudobulges using HST archival images in the optical
wavelength regime ($B$ through $I$) . This makes bulge classification
subject to the effects of dust. However, the structures used to
identify pseudobulges are usually experiencing enhanced star formation
rates, and are easier to detect in the optical region of the spectrum
where the mass-to-light ratios are more affected by young stellar
populations, rather than in the near infrared where the effects of
dust are lesser. Classical bulges may have dust in their center, as do
many elliptical galaxies \citep{Laueretal05}. The presence of dust
alone is not enough to classify a galaxy as containing a pseudobulge.

Another caveat when using morphology is that the structures we wish to
identify as well as the color of a galaxy can be distorted or altered
during early stages of a merger. For example NGC~3169 and NGC~3166
form a close pair (separation is $\sim50$~kpc). \cite{sandage94} note
that NGC~3169 shows a warped disk and peculiar morphology. Both of
these galaxies have nuclear structure that is similar to structure
found in pseudobulges.  However, given the presence of the companion,
we cannot say whether the central structure represents a
secularly-evolved pseudobulge or is due to short-term merger-induced
gas inflow and star formation (see \citealp{KJB04}).

We use the NASA Extragalactic Database (NED) to search for any
evidence of close companions of similar magnitude, tidal distortions,
or peculiar morphology. We remove those galaxies which seem to be
interacting with other galaxies from our sample. Three galaxies in our
sample have companions at $\sim 100$~kpc, which do not appear to
affect the morphology of these galaxies' disks.

Of the 39 galaxies in our sample, 10 galaxies are classified as having
a classical bulge (3 S0, 3 Sa, and 4 Sab) and 29 galaxies are
classified as having a pseudobulge (3 S0, 4 Sa, 1 Sab, 5 Sb, 6 Sbc,
and 10 Sc). We do not distinguish between barred and unbarred galaxies
in this work. The sample ends up favoring pseudobulges over classical
bulges, most likely due to the constraint of even sampling along
Hubble types as pseudobulges are more common in late type galaxies.

Table~1 lists the galaxies in our bulge sample along with
their bulge classification, Hubble types, magnitudes, colors, and
other derived quantities (described below).

\subsection{Photometry}\label{sec:photometry}

We calculate total magnitudes and colors from direct integration of 2D
surface brightness profiles in the SDSS $u$ and $r$ band images. We
use the code of \cite{BDM98} on images available
publicly from the SDSS archive \citep{SDSS-DR4}.

First, interfering foreground objects are identified in each image and
masked manually. Then, isophotes are sampled by 256 points equally
spaced in an angle $\theta$ relating to polar angle by $\tan \theta =
a/b\,\tan \phi$, where $\phi$ is the polar angle and $b / a$ is the
axial ratio.  An ellipse is then fitted to each isophote by least
squares. The software determines six parameters for each ellipse:
relative surface brightness, center position, major and
minor axis lengths, and position angle along the major axis.

To calculate the structural quantities central surface brightness,
$\mu_0$, and global S\'ersic index, $n$, we fit a S\'ersic function,
\begin{equation}\label{eq:sersic-fit}
  \mu(r)=\mu_0+(r/r_0)^{1/n},
\end{equation}
to the mean isophote axis of SDSS surface brightness profiles. It is
well known that surface brightness profiles of intermediate type
galaxies are not well described by single component models. At least
two component functions (bulge plus outer disk) are required; also,
many galaxies contain bars, rings and ovals that further complicate
the surface brightness profile. For the nearby galaxies in our bulge
sample a single-component S\'ersic fit is clearly not an excellent fit
to the galaxies' light profiles. However, we wish to compare these
fits to the published manifestations of the galaxy
bimodality. Therefore, we must calculate quantities similar to those
in large surveys. Typical root-mean-square deviations of our S\'ersic
fits to the galaxy profiles are 0.1-0.2 mag arcsec$^{-2}$.

Bulge-to-total ($B/T$) ratios are calculated by fitting S\'ersic
functions combined with an exponential for the outer disk to the 1D
surface brightness profile of each galaxy:
\begin{equation}\label{eq:bt-fit}
  \mu(r)=\mu_{0,b}+\left(\frac{r}{r_{0,b}}\right)^{1/n_b} \,\ + \,\
  \mu_{0,d} + \frac{r}{h} \ ,
\end{equation}
where the $\mu_{0,b}$ and $r_{0,b}$ reflect central surface brightness
and scale length of the bulge, while $\mu_{0,d}$ and $h$ are the
analogous quantities for the outer disk. $n_b$ is the S\'ersic index
of the bulge.

Equation \ref{eq:bt-fit} is fit to surface brightness profiles
generated from HST archival images combined with SDSS $r$ band
photometry as discussed above. For calculating $B/T$, we allow the
bulge and the disk component to have individual ellipticities, which
we take to be the average ellipticity within each component. This
definitely adds a little uncertainty to the resulting $B/T$, as both
bulges and disks are known to have varying ellipticity profiles
\citep{FP03}.  However, for our purposes this method is
sufficient. Finally, the $B/T$ that we quote is the ratio of the
radially integrated S\'ersic function and the radially integrated sum
of the S\'ersic and the exponential. We have checked that there is no
trend of $B/T$ with inclination for pseudobulges, classical bulges, or
the combined sample.

We refer the reader to Table~1, where these quantities
are listed for all our bulge galaxies.


\section{Dependence of bimodal galaxy properties on bulge
  type}\label{sec:bulge-type}

\begin{figure*}[t]
  \centering
  \includegraphics[width=\textwidth]{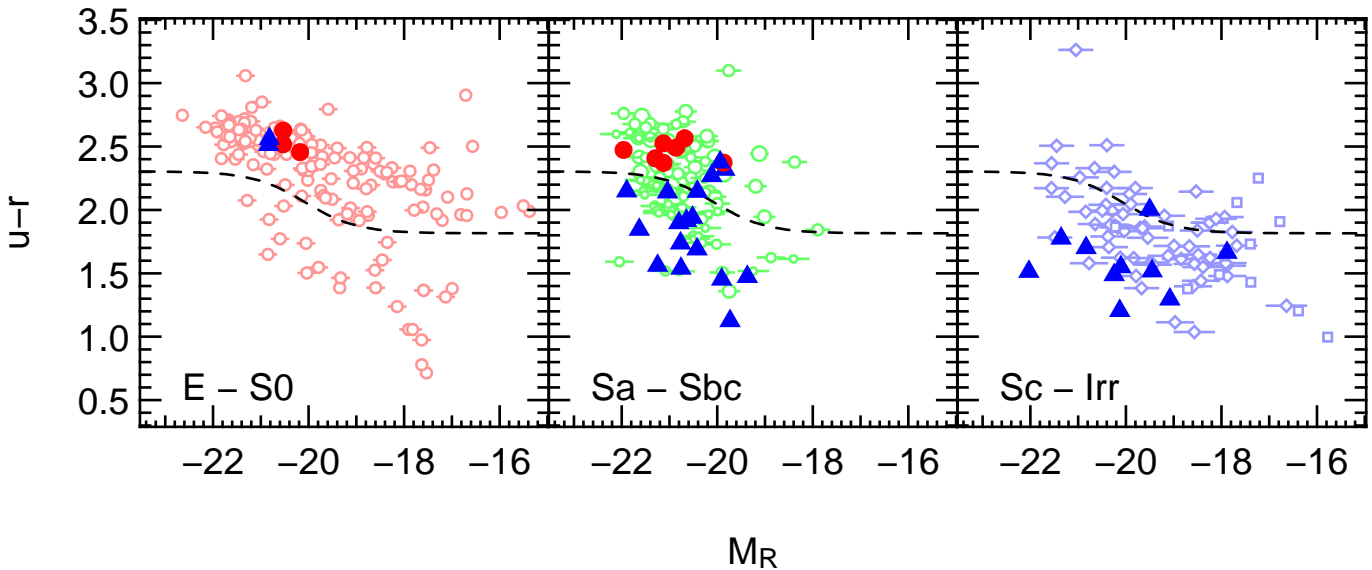}
  \caption{The location of three different galaxy populations is shown
    in global color vs.\ total magnitude space, from left to right:
    early-type (E-S0), intermediate-type (Sa-Sbc), and late-type
    (Sc-Irr). Galaxies identified as having pseudobulges are
    represented by filled triangles, galaxies with classical bulges
    are shown as filled circles. Galaxies without bulge identification
    are shown as open symbols for comparison. The dashed line
    separates the red sequence from the blue cloud following
    \citet{Baldryetal04}.}\label{fig:cm}
\end{figure*}

Does the shift from the blue cloud to the red sequence coincide with a
shift in predominance of the disk to the bulge?

Using extensive 2D photometry, \citet{Driveretal06} find that the
bimodality of galaxies in color--magnitude space becomes more
pronounced when the color is constrained to the central region of the
galaxy. They also show that the concentration of the galaxy,
parameterized by global S\'ersic index, participates in the bimodality
in galaxy properties as well. This motivates them to suggest that the
fundamental division of galaxies is merely that of bulges (old, red,
high concentration) and disks (star forming, blue, diffuse). With this
two-component nature of galaxies, the position of an object on the
blue--red divide is thought of as a function of the bulge-to-total
ratio, $B/T$, alone.

The existence of intermediate-type Sa-Sbc galaxies on both the red and
blue sequence challenges the suggestion that the bimodality of
galaxies divides along the disk-bulge dichotomy. Intermediate type
galaxies have roughly constant (and large) bulge-to-total ratios on
average \citep{SdV86}. Bulge-to-total ratios begin to fall much more
rapidly only at Hubble types Sbc-Sc and later. Moreover, disk color
and bulge color are correlated: redder bulges reside in redder disks
\citep{PB96,deJong96a,MacArthur04}. Galaxies are not made by randomly
mixing disk and bulge components.

At issue, therefore, is whether the galaxy bimodality is just a
product of changing emphasis of each subcomponent (i.e.\ simply
$B/T$), or possibly a signature of differing evolutionary histories of
the whole galaxy.

The existence of a dichotomy among bulges themselves offers the
opportunity to test this. The question becomes whether at intermediate
Hubble types of Sa-Sbc, where both bulge types are found, the color
bimodality separates galaxies at some bulge-to-total ratio, or,
whether it separates galaxies of different bulge type, irrespective of
bulge-to-total ratio.

\subsection{The color -- magnitude plane and Hubble types}

We examine galaxies of Hubble types spanning S0-Sc in the global color
($u\!-\!r$) versus total magnitude ($M_r$) plane, and we mark them
according to their bulge type. Fig.~\ref{fig:cm} shows the location of
galaxies with classical bulges (round symbols) and galaxies with
pseudobulges (triangles; identified by bulge morphology as discussed
in Sect.~\ref{sec:ident-pseudo}) in our sample in the $u\!-\!r$ vs.\
$M_r$ plane. Note that we plot the total galaxy color and total
magnitude, not the bulge color and magnitude. We merely label the
galaxies by their bulge type.

As a reference sample, we also plot 542 galaxies selected from the
intersection of the SDSS-DR4 spectroscopic catalog and the RC3, having
inclination $i < 35\degr$, and that are at a distance $z <
0.02$. These galaxies divide into Hubble types as follows: 50 E, 112
S0, 48 Sa, 36 Sab, 67 Sb, 57 Sbc, 52 Sc, 40 Scd, 47 Sd, 17 Sm and 16
Irr. We use SDSS redshifts for distances and SDSS ``model'' magnitudes
for colors and total magnitudes for these objects.

We note here that the magnitudes of the galaxies that we classify as
having classical bulges or pseudobulges are computed by our own
ellipse fitting discussed in Sect.~\ref{sec:photometry}. This may give
different results compared to the SDSS model magnitudes which we use
only for the reference sample objects in the color--magnitude
plane. However, for galaxies with low total S\'ersic index (as the
intermediate types mainly are) there is very little difference in the
type of magnitude \citep{Grahametal05}. Also note that we do not
correct the colors and magnitudes for the effects of extinction by
dust. This may cause some disk galaxies to appear redder than their
stellar populations are. We moderate this effect by restricting
ourselves to low-inclination galaxies. However, some disk galaxies at
the locus of the red sequence may have been moved there by the effect
of dust.  For illustrative reasons, we also plot the line dividing
blue from red galaxies following \citet{Baldryetal04}.

{\bf Late types} (Right panel in Fig.~\ref{fig:cm}). As has been shown,
late type galaxies (type Sc and later) are almost entirely on the blue
sequence \citep[e.g.][]{Stratevaetal01}. Note the caveat on dust
extinction in disk galaxies discussed above; the reddest galaxies in
this bin are most likely affected by dust extinction. We emphasize
that the panel with Sc-Irr galaxies does not contain a single
classical bulge. As the Hubble sequence progresses toward later types,
galaxies tend to have small bulges or no bulge at all. This is
indicative of a less violent past, as it is very likely that these
galaxies have not experienced a merger event that would have formed a
(classical) bulge since the time of formation of their disks. The fact
that these galaxies seem to contain pseudobulges if they have a bulge
at all, provides a strong reinforcement of this statement.

{\bf Intermediate types} (Middle panel in Fig.~\ref{fig:cm}). The
intermediate type Sa-Sbc galaxies give us a sample on which to test
our hypothesis. Nearly all (87\%) galaxies with pseudobulges are bluer
than the red--blue divide, while all the galaxies with classical
bulges are redder than the divide.  To show that this is not simply
the consequence of the pseudobulge galaxies having lower
bulge-to-total ratios than the classical bulge galaxies, recall first
that the number of pseudobulges in our sample is not a step function
at some late Hubble type (5 Sa, 2 Sab, 4 Sb, 6 Sbc) and that as noted
above, at these intermediate types the Hubble sequence is not a strong
sequence of bulge-to-total ratios (see \citealp{SdV86} and the review
by \citealp{RH94}).

\begin{figure}[t]
  \includegraphics[width=8cm]{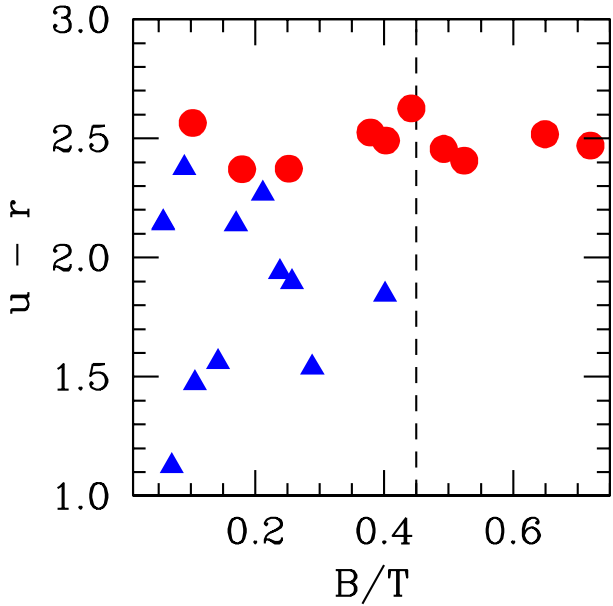}
  \caption{The distribution of bulge-to-total ratios, $B/T$, of
    intermediate type (Sa-Sbc) galaxies with pseudobulges (blue
    triangles) and classical bulges (red filled circles) with respect
    to their global $u\!-\!r$ color. The dashed line marks $B/T =
    0.45$. }\label{fig:bt}
\end{figure}

In fact, in the range of $B/T$ values spanned by galaxies with
pseudobulges we find plenty of classical bulge galaxies as well.  This
is illustrated in Fig.~\ref{fig:bt}, where we plot global $u\!-\!r$
against $B/T$ for 24 of our bulge galaxies. We calculate $B/T$ for all
10 classical bulge galaxies and all the 16 intermediate-type
pseudobulge galaxies. Two of the latter (NGC~1068 and NGC~5691) are
not well-fit by a two-component model (Equation~\ref{eq:bt-fit}) and
hence we cannot obtain reliable $B/T$ numbers for these and they are
not included in the plot (see also Table~1). Therefore the plot
contains 14 pseudobulge galaxies.

In our sample, galaxies on the red sequence with classical bulges have
$B/T$ ratios as low as 10\%. Galaxies with pseudobulges have $B/T$
ratios as high as 40\%. The majority of galaxies with classical bulges
in our (small) sample have $B/T$ values in the same range as the
galaxies with pseudobulges. Even at the lowest $B/T \sim 0.1$ values
in our intermediate type Sa-Sbc galaxies, the assignment of a galaxy
to the red sequence or the blue cloud is predicted by its bulge type.

It is true that classical bulge galaxies extend to greater $B/T$
values than do pseudobulge galaxies. This is easily understood given
the different formation channels that are thought to lead to classical
bulges and to pseudobulges. Classical bulges are an extension of
elliptical galaxies (formed via mergers) that happen to have a disk
around them. This sort of evolution naturally extends all the way to
$B/T = 1$ (i.e.\ a pure elliptical galaxy; see also
\citealp{KB96}). Pseudobulges form secularly by rearranging disk
material. Therefore it seems unlikely that a disk would be able to
make a pseudobulge equal in size to itself ($B/T \simeq 0.5$) through
secular evolution.

Also note that in Fig.~\ref{fig:cm}, there is no significant
difference in the range of absolute magnitudes spanned by the
pseudobulge galaxies and that of the classical-bulge
galaxies. Pseudobulge galaxies are not systematically fainter in
$M_r$.

Concluding this discussion, we find that the red--blue bimodality
cannot be a function of decreasing bulge prominence alone. Our results
show that it is a function of bulge type. Pseudobulges are in blue
galaxies and classical bulges are in red galaxies. Furthermore,
galaxies with pseudobulges behave just like pure disk galaxies if we
compare their distribution in global color to the distribution of pure
disk (late-type) galaxies in Fig.~\ref{fig:cm}. The type of bulge a
galaxy has is a signpost for an evolutionary history of the total
galaxy.

{\bf Early types} (Left panel in Fig.~\ref{fig:cm}). The early-type
bin (E - S0) is almost entirely populated by red sequence
galaxies. There are three galaxies that we identify as harboring
pseudobulges in this panel. All three peudobulges are in S0 galaxies
and these are on the red sequence. This illustrates a caveat when
dealing with pseudobulges. They do not have to be young. A disk galaxy
could have undergone secular evolution long ago and formed a
pseudobulge. This is well illustrated by the aforementioned
correlation between disk color and bulge color. As a side note, this
implies that identifying pseudobulges using color only is bound to
misclassify some pseudobulges and underestimate their number.  Also,
S0 galaxies exist at a wide range of bulge-to-total ratios, a point
noticed by \citet{vdB76}. The processes that are thought to make S0
galaxies (e.g.\ gas stripping by ram pressure, harassment;
\citealp{MKLDO96}) operate independently of the processes that make
bulges.  It is reasonable to believe that the evolution which makes a
galaxy an S0 happens independently of the secular evolution that makes
a pseudobulge (see the discussion in \citealp{KK04}). Therefore the
position of S0 galaxies in color--magnitude space may be due to
separate phenomenona, rather than posing a counter example to our
hypothesis.

\subsection{Global galaxy structure}

\begin{figure*}[t]
  \centering
  \includegraphics[width=14cm]{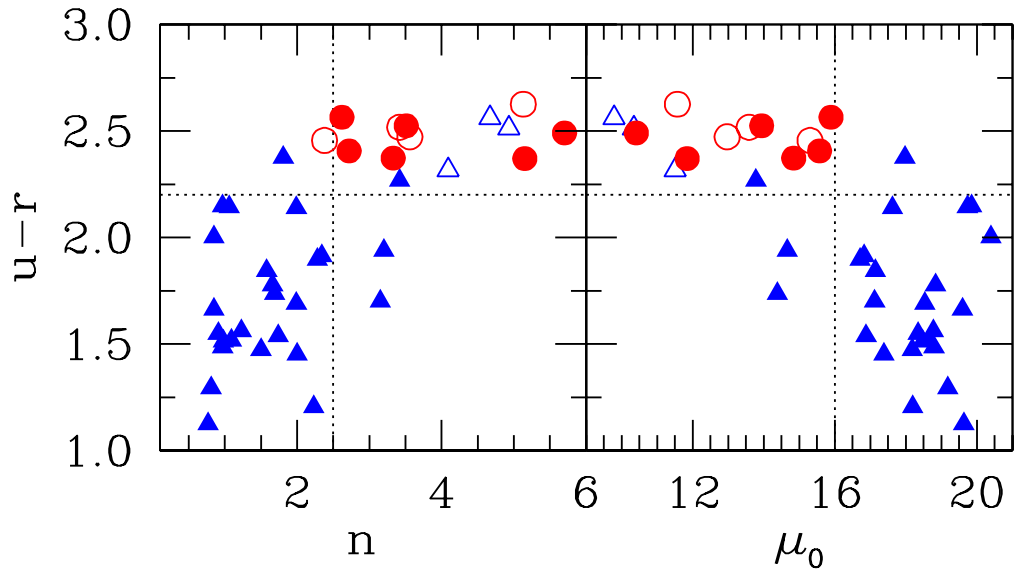}
  \caption{Dependence of global structure on bulge type. The left
    panel shows the S\'ersic index from a fit of a single S\'ersic law
    (Eq.~\ref{eq:sersic-fit}) to the surface brightness profile of the
    whole galaxy, plotted against global color. The right panel shows
    the corresponding central surface brightness versus color.  Blue
    triangles represent pseudobulges, red circles represent classical
    bulges; S0 galaxies are plotted as open symbols. In both panels
    the horizontal line denotes $u!\-\!r=2.2$. The vertical line in
    the left panel denotes $\mu_0=16$, and the horizontal line in the
    right panel denotes $n=2.5$.}\label{fig:struct}
\end{figure*}

\cite{Blantonetal03} show that the relation of structure to galaxy
color is markedly different for red and blue galaxies. This can be
illustrated by plotting the S\'ersic index and a characteristic
surface brightness against color. In Fig.~\ref{fig:struct}, we
illustrate the dependence of galaxy structure on bulge type. We mark
the dividing line of red and blue galaxies with a horizontal line at
$u\ -\ g=2.2$ \citep{Stratevaetal01}. Open symbols represent S0
galaxies, which likely arise due to distinct phenomena not related to
bulge formation as discussed above, and thus do not follow the normal
behavior for pseudobulge galaxies. In general, blue galaxies are more
diffuse and have lower S\'ersic index than galaxies with classical
bulges. In both structural parameters there is a sharp transition from
galaxies with pseudobulges to galaxies with classical bulges. Again,
as in Fig.~\ref{fig:cm}, we plot total color and S\'ersic parameters
from a global single-component fit. We mark galaxies by the type of
their bulge.

The surface brightness at zero radius is recovered from the single
component S\'ersic fits (Eq.~\ref{eq:sersic-fit}). Note that we report
the fit value, which should be taken as a characteristic value of the
galaxy as a whole. It is also worth noting that central surface
brightness is more sensitive to the dichotomy in bulge types than
surface brightness at the effective radius. Galaxies with classical
bulges form a tight sequence in color that begins at $\mu_{0,r} \sim
16$ (vertical line in the left panel) and extends to higher central
surface densities, well populated by giant elliptical galaxies. There
is a sharp transition at $\mu_{0,r} \sim 16$.  Below this surface
density, the diagram becomes completely populated by galaxies with
pseudobulges.

The S\'ersic index is normally interpreted as a parametrization of the
shape of a surface brightness profile of a galaxy.  A typical disk
galaxy has $n = 1$ and an intermediately sized elliptical galaxy has
$n \sim 4$. In the right panel of Fig.~\ref{fig:struct}, galaxies with
pseudobulges have lower global S\'ersic index than galaxies with
classical bulges. And galaxies with classical bulges do not have
global S\'ersic index smaller than $n \sim 2.5$ (vertical line in the
right panel).

Galaxies with pseudobulges populate a broader range of color, yet the
majority of pseudobulge galaxies are restricted to a more narrow range
in central surface brightness than classical bulges. The smaller
variation in $\mu_0$ and exponential surface brightness profiles are
well known properties of disk galaxies \citep{freemanlaw}. We restate
that our galaxies with pseudobulges and galaxies with classical bulges
both populate the intermediate Hubble types Sa-Sbc. Thus, at the same
Hubble type, galaxies with pseudobulges are more like pure disk
galaxies than galaxies with classical bulges.

What is compelling is not only that global S\'ersic index and central
surface density are lower. In fact, the distribution of central
surface brightness alone is not dichotomous. It is the combination of
these quantities that reveals the structural dichotomies. The
distribution in both the color -- S\'ersic and the color -- $\mu_0$
plane is completely different for galaxies with pseudobulges and
classical bulges. Further, the transition in this behavior is
coincident with the transition from red galaxies to blue.

The behavior of color and structure is definitely more regular among
classical bulges. This makes sense in context of the formation picture
of these two types of bulges. Pseudobulges are structures that are
continuously evolving. Therefore when we look at a sample of
pseudobulge galaxies we are seeing them in many different stages of
evolution. As the pseudobulge grows galaxy properties such as color,
$\mu_0$ and $n$ will certainly change.  However, classical bulges are
thought to form in one or multiple discrete and short-lived events
separated by periods of passive evolution. If the young stellar
populations made in these events age into old, red populations on a
timescale that is shorter than the time between star formation events,
then classical bulges will spend most of their time looking
homogeneously red. We find that the galaxies harboring these classical
bulges are globally red. This implies that since the formation of the
classical bulge, there has been relatively little evolution in the
galaxy that contains it as a whole.  Thus, properties of galaxies with
classical bulges show little scatter compared to pseudobulge
galaxies. Studies considering bulges as a heterogeneous class will
likely shed light on such differences.

\begin{deluxetable*}{llclclllll}
\tablewidth{0pt}
\tablecaption{Global colors and structural data for the bulge sample}
\tablehead{\colhead{Identifier} & \colhead{Type} & \colhead{Inclination} &
  \colhead{$m-M$} & \colhead{Bulge\tablenotemark{(a)}} &
  \colhead{$u\!-\!r$} & \colhead{$M_r$} &
  \colhead{$\mu_0$\tablenotemark{(b)}} & \colhead{$n$\tablenotemark{(b)}} &
  \colhead{$B/T$\tablenotemark{(c)}} \\
  \colhead{} & \colhead{RC3} & \colhead{degrees} & \colhead{mag} & \colhead{} &
  \colhead{mag} & \colhead{mag} & \colhead{mag/arcsec$^2$} &
  \colhead{global S\'ersic} & \colhead{}}
\startdata
 NGC~2639     & Sa  & 44 & 33.24 & c & 2.47 & -21.96 & 12.97 & 3.56 & 0.72    \\
 NGC~2775     & Sab & 40 & 30.95 & c & 2.41 & -21.29 & 15.57 & 2.72 & 0.52    \\
 NGC~2880\tablenotemark{(d)}
              & S0  & 68 & 31.83 & c & 2.52 & -20.53 & 13.59 & 3.42 & 0.65    \\
 NGC~2962\tablenotemark{(d)}
              & S0  & 67 & 32.12 & c & 2.63 & -20.53 & 11.57 & 5.13 & 0.44    \\
 NGC~3031     & Sab & 59 & 27.63 & c & 2.49 & -20.86 & 10.41 & 5.70 & 0.40    \\
 NGC~3898     & Sab & 57 & 31.49 & c & 2.53 & -21.13 & 13.94 & 3.51 & 0.38    \\
 NGC~4379     & S0  & 43 & 31.59 & c & 2.46 & -20.18 & 15.30 & 2.38 & 0.49    \\
 NGC~4698     & Sab & 51 & 31.59 & c & 2.37 & -21.13 & 11.84 & 5.15 & 0.18    \\
 NGC~4772     & Sa  & 64 & 30.64 & c & 2.37 & -19.87 & 14.84 & 3.33 & 0.25    \\
 NGC~5448     & Sa  & 64 & 32.34 & c & 2.57 & -20.68 & 15.89 & 2.62 & 0.10    \\ \hline

 NGC~1068     & Sb  & 21 & 30.46 & p & 1.74 & -20.77 & 14.38 & 1.69 & ---\tablenotemark{(e)} \\
 NGC~1084     & Sc  & 46 & 30.56 & p & 1.55 & -20.10 & 18.34 & 0.91 & \ldots  \\
 NGC~2681     & S0  & 0  & 30.20 & p & 2.32 & -19.84 & 11.51 & 4.09 & \ldots  \\
 NGC~2782     & Sa  & 49 & 32.70 & p & 1.56 & -21.25 & 18.77 & 1.23 & 0.14    \\
 NGC~2859     & S0  & 33 & 31.78 & p & 2.51 & -20.83 & 10.34 & 4.93 & \ldots  \\
 NGC~2950     & S0  & 62 & 31.49 & p & 2.56 & -20.82 & 9.79  & 4.67 & \ldots  \\
 NGC~2976     & Sc  & 61 & 27.63 & p & 1.66 & -17.88 & 19.59 & 0.85 & \ldots  \\
 NGC~3259     & Sbc & 61 & 31.97 & p & 1.47 & -19.37 & 18.18 & 1.50 & 0.11    \\
 NGC~3338     & Sc  & 54 & 32.06 & p & 1.78 & -21.36 & 18.83 & 1.66 & \dots   \\
 NGC~3351     & Sb  & 42 & 29.24 & p & 2.38 & -19.94 & 17.97 & 1.81 & 0.09    \\
 NGC~3359     & Sc  & 53 & 30.98 & p & 1.21 & -20.13 & 18.19 & 2.23 & \ldots  \\
 NGC~3368     & Sab & 55 & 29.24 & p & 2.27 & -20.12 & 13.78 & 3.42 & 0.21    \\
 NGC~3627     & Sb  & 57 & 29.17 & p & 1.94 & -20.52 & 14.66 & 3.20 & 0.24    \\
 NGC~3642     & Sbc & 32 & 32.09 & p & 1.92 & -20.64 & 16.82 & 2.34 & 0.13    \\
 NGC~3810     & Sc  & 48 & 29.80 & p & 1.52 & -19.45 & 18.65 & 1.09 & \ldots  \\
 NGC~4030     & Sbc & 40 & 31.70 & p & 1.85 & -21.63 & 17.14 & 1.58 & 0.40    \\
 NGC~4051     & Sbc & 36 & 30.74 & p & 1.69 & -20.42 & 18.53 & 1.99 & 0.07    \\
 NGC~4123     & Sc  & 48 & 30.91 & p & 2.00 & -19.50 & 20.39 & 0.85 & \ldots  \\
 NGC~4152     & Sc  & 40 & 32.31 & p & 1.49 & -20.25 & 18.79 & 0.97 & \ldots  \\
 NGC~4254     & Sc  & 32 & 31.59 & p & 1.51 & -22.03 & 18.47 & 0.98 & \ldots  \\
 NGC~4380     & Sb  & 59 & 31.59 & p & 2.15 & -20.42 & 19.73 & 1.06 & 0.06    \\
 NGC~4384     & Sa  & 42 & 32.60 & p & 1.13 & -19.73 & 19.63 & 0.77 & 0.07    \\
 NGC~4500     & Sa  & 50 & 33.18 & p & 1.54 & -20.76 & 16.88 & 1.74 & 0.29    \\
 NGC~4536     & Sbc & 59 & 32.02 & p & 2.15 & -21.90 & 19.85 & 0.97 & 0.06    \\
 NGC~4647     & Sc  & 38 & 31.59 & p & 1.70 & -20.83 & 17.12 & 3.15 & \ldots  \\
 NGC~4900     & Sc  & 19 & 30.41 & p & 1.30 & -19.08 & 19.18 & 0.81 & \ldots  \\
 NGC~5055     & Sbc & 56 & 29.21 & p & 1.90 & -20.80 & 16.71 & 2.28 & 0.26    \\
 NGC~5691     & Sa  & 42 & 31.97 & p & 1.45 & -19.91 & 17.38 & 2.00 & ---\tablenotemark{(e)} \\
 NGC~5806     & Sb  & 60 & 32.05 & p & 2.14 & -21.04 & 17.62 & 1.99 & 0.17    \\
\enddata
\tablenotetext{(a)}{p -- pseudobulge; c -- classical bulge.}
\tablenotetext{(b)}{global S\'ersic index and central surface
  brightness determined by a fit of Eq.~1 to the profiles of all
  classical bulge galaxies and the intermediate-type pseudobulge
  galaxies.}
\tablenotetext{(c)}{$B/T$ determined by a fit of Eq.~2 to the profile.}
\tablenotetext{(d)}{Erwin (2004) finds nuclear bar or inner disk, we do not.}
\tablenotetext{(e)}{These galaxies are not well-fit by a two-component model.}
\end{deluxetable*}
\nocite{Erwin04}


\section{Summary and Discussion}\label{sec:summary}

We examine galaxies of Hubble types spanning S0-Sc in the global color
versus magnitude plane, marking them according to their bulge type. We
classify them as having pseudobulges or classical bulges by analyzing
the morphology of the bulge using HST imaging.

We show that the type of bulge a galaxy has is a good predictor of
where that galaxy will fall in the red-blue and structural galaxy
dichotomies. Galaxies with pseudobulges lie in the blue cloud. They
have the same global color as galaxies with no bulge at all (pure disk
galaxies).  On the other hand, galaxies having a classical bulge (and
elliptical galaxies) lie on the red sequence. We have further shown
that this is not an effect of lower bulge-to-total ratios in
pseudobulge galaxies.  Additionally, we show that galaxies with
pseudobulges have lower global S\'ersic index and lower central
surface density than galaxies with classical bulges.

Our results imply that the processes involved in the formation of
galactic bulges are coupled to the processes that form the entire
galaxy. Galactic disks and classical bulges remain the two fundamental
structural components of galaxies, yet their relative prominence is
not what determines the location of an (intermediate-type) galaxy with
respect to the color and structural bimodalities. It is the presence
of a classical bulge and the influence this bulge has and its
formation had on the galaxy that places the galaxy on the red sequence
today. Another way of putting this is to say that increasing the
bulge-to-total ratio of an intermediate-type galaxy with a pseudobulge
will not automatically move the galaxy to the red sequence, and
likewise, a galaxy with a classical bulge is on the red sequence no
matter how small that bulge is.

Thus, the location of a galaxy with respect to the bimodality
does -- at least in part -- reflect differing evolutionary paths of
the whole galactic system. It is not merely a reflection of different
emphasis of the disk and bulge subcomponents.

We wish to reiterate that interpreting the red-blue bimodality as due
to the fundamental distinction between disks and classical bulges is
not necessarily incorrect. However what is incorrect is to say that
the bimodality is merely the linear combination of these two
components that determines the location of a galaxy with respect to
the bimodality. Also, if one defines the components by photometric
decomposition only as has been common practice in the literature, then
one is likely to obtain ambiguous results.

As reviewed in \citet{KK04} and outlined in
Sect.~\ref{sec:introduction}, a pseudobulge forms through dynamical
processes inside the disk. We also know that the existence of a thin
disk signals that the galaxy has not undergone a major merger since
the formation of the disk \citep{TO92}. The exact mass ratio in a
merger event that is needed to destroy a disk is still under debate
\citep{VW99,HC06}, however, it is believed to be of the order of 0.2.
All merger events above this mass ratio are believed to result in
dynamically hot stellar systems.  This leads to the well-known problem
of forming late-type disks in cosmological simulations; the disks in
these simulations suffer too many mergers and thicken or are destroyed
\citep{steinmetz2002}. The problem of forming these galaxies in
cosmological simulations gets much worse when we realize that galaxies
with pseudobulges should be included with pure disk galaxies in this
sense: they have not suffered a major merger event since the formation
of the disk.

The processes that are believed to make pseudobulges require not only
a disk to be present, but that disk must be dynamically very cold in
order to be unstable to non-axisymmetric instabilities such as bars
that are needed to rearrange gas and stars to form a pseudobulge (see
\citealp{KK04} for a detailed review). Thus, pseudobulges, like thin
disks, are signatures of a quiet merger history. Pseudobulge formation
timescales have been estimated as $\sim 1$~Gyr \citep{KK04,Fisher06}
based on current star formation rates. Now consider the time necessary
to form a bar and build up a sufficiently high nuclear gas density to
form stars. It is quite plausible that these galaxies have remained
cold (and thus free of significant mergers) since the formation of
their disks.

How can we explain our result that classical-bulge galaxies lie
exclusively on the red sequence? Classical bulge galaxies are thought
to have undergone violent relaxation, most likely during a major
merger in their past, presumably one involving lower-mass galaxies to
form a typical bulge of today's intermediate type Sa-Sbc
galaxies. Intermediate mass mergers happen predominantly in the
earlier universe, when the halos harboring these objects
assemble. Today, the surviving halos of this mass scale are
incorporated into bigger structures. For example, the large and small
Magellanic Clouds are sub-halos of the Milky Way system.  At early
times, intermediate-mass mergers frequently occur, and there is still
enough gas left to be accreted, so that a disk forms around the newly
formed bulge. The result is a disk galaxy that contains a classical
bulge.

As the merger rate was much higher at early times in the universe, it
is less likely that a pure disk galaxy would have survived this epoch
\citep{steinmetz2002}. Therefore old, red galaxies of Hubble type Sc
and later do not commonly exist today. Also, the only way to grow a
pseudobulge is to have a relatively long period of disk evolution (of
the order of a few Gyr) undisturbed by mergers. This is much more
likely to happen in low-density environments at later epochs, possibly
$z < 1$.  Therefore, it is natural that pseudobulges live in blue,
relatively younger galactic systems that are more typical of
low-density environments today.

Parenthetically, in high-density environments a few objects may evolve
toward the red sequence within $\sim 1$~Gyr \citep{Blanton06} by gas
stripping and quenching of their star formation irrespective of their
merger history and resulting bulge type, most likely to become S0s.

The environmental dependence of galaxy colors is well established by
studies that link galaxy properties to the environment in which they
are found.  We first recall the morphology-density relation
\citep{dressler1980}, namely that early-type (and thus red) galaxies
are more clustered than late-type (and thus blue)
galaxies. \cite{li2006} restate this in the modern context, showing
that not only color, but also surface density, concentration, and
4000\AA\ break strength all correlate with environmental density in
the same way (see also \citealp{blanton2005environ}). Since we find
all our (non-S0) pseudobulges in blue-cloud galaxies which are
preferentially found in low-density environment, we take this as a
reinforcement of the conclusion that pseudobulges indicate galaxies
with quiescent merger histories.

Evidence for a quiet merger history of blue galaxies is also given by
\cite{Blanton06}. He finds that the number density of blue-sequence
galaxies does not change significantly from $z\sim1$ to $z\sim0.1$.
It seems likely that blue sequence is experiencing a quiescent form of
evolution over the recent epochs ($z\lesssim 1$), and therefore the
blue (disk) galaxies have had time to form pseudobulges. We are
compelled to point out an important caveat for interpreting any high
redshift results on galaxy evolution. Blue galaxies at high redshifts
are not likely to be replicas of today's blue sequence galaxies. It is
unfortunate that pseudobulge detection requires such high resolution
techniques, otherwise we could test our hypothesis at higher
redshifts.  The evolution in blue galaxies beyond $z\sim 1$
\citep{gerke2006} is possibly due to separate phenomenon from what
shapes todays blue galaxies (e.g.\ the movement of galaxies from the
blue sequence to the red). However, the result that the blue sequence
appears to be less violent over the past few billion years appears
established.

This leads to a picture where a classical bulge is indicative of a
halo formation and assembly history of the galaxy that occurred
earlier, leading to older stars in the bulge and in the disk, was more
violent, and ultimately the system is red today. On the contrary, a
pseudobulge, even at the same Hubble type, is indicative of a quiet
assembly history that is much more likely at later times and therefore
also leads to bluer and younger systems (See \citealp{GKK01} for the
dependence of halo merger histories on environment).  Thus, the
presence of a classical bulge or the presence of a pseudobulge (or no
bulge at all) are indicators of different evolutionary histories of
the whole galaxy.

Along another line of thought it has suggested that feedback from
accreting black holes might be responsible for quenching the star
formation history of elliptical galaxies and classical bulges and
thereby contribute to the formation of the red sequence
(\citealp[e.g.][]{SdMH05,SSB05,deLuciaetal06,Boweretal06}). We know
that some pseudobulges contain supermassive black holes since they
contain AGN (for example NGC~1068, NGC~3368, and
NGC~4051). Furthermore, there is compelling evidence that these lie on
the $M_{BH}$--$\sigma$ relation \citep{KG01}. If this turned out to be
true, we need to explain why pseudobulges and their supermassive black
holes obey the same $M_{BH}$--$\sigma$ relation as classical bulges
and ellipticals do despite their disk-like properties, and at the same
time why feedback was not able to quench their star formation in spite
of doing just the same in classical bulges of the same size. We do not
know of convincing answers to these two questions. The latter might
imply that AGN feedback is not important in shaping the star formation
history of low-mass systems, or, it might be related to different
formation timescales of the mergers that form classical bulges and the
secular evolution processes that are thought to lead to pseudobulge
formation. In fact, \citet{Fillipenko03} find an AGN that obeys the in
$M_{BH}$--$\sigma$ relation in NGC~4395, a low-mass galaxy that does
not contain any bulge, classical or pseudo \citealp[see
also][]{GH06,GBH06}.

We take our result as clear observational evidence that initial
environmental conditions are responsible for many global properties of
a galaxy.  The merging history of a galaxy is ultimately the driver
behind the observed bimodality in the structural, stellar population,
and kinematic properties of a galaxy.  Bulge types are a signpost of
merger history and because of this they predict the position of the
whole galaxy with respect to the bimodality in color and structural
properties.


\acknowledgments

DBF wishes to thank A.~Filippenko and the University of California at
Berkeley for providing support. ND and DBF thank the Max-Planck
Society for support during this project. We also thank J.~Kormendy for
stimulating and valuable discussions, and the anonymous referee for
useful comments.

This research was supported by the National Science Foundation under
grant AST 06-07490.

Some of the data presented in this paper were obtained from the
Multi-mission Archive at the Space Telescope Science Institute
(MAST). STScI is operated by the Association of Universities for
Research in Astronomy, Inc., under NASA contract NAS5-26555. Support
for MAST for non-HST data is provided by the NASA Office of Space
Science via grant NAG5-7584 and by other grants and contracts.

This research has made use of the NASA/IPAC Extragalactic Database
(NED) which is operated by the Jet Propulsion Laboratory, California
Institute of Technology, under contract with the National Aeronautics
and Space Administration.

Funding for the SDSS and SDSS-II has been provided by the Alfred P.\
Sloan Foundation, the Participating Institutions, the National Science
Foundation, the U.S.\ Department of Energy, the National Aeronautics
and Space Administration, the Japanese Monbukagakusho, the Max Planck
Society, and the Higher Education Funding Council for England. The
SDSS Web Site is {\tt http://www.sdss.org/}.

The SDSS is managed by the Astrophysical Research Consortium for the
Participating Institutions. The Participating Institutions are the
American Museum of Natural History, Astrophysical Institute Potsdam,
University of Basel, Cambridge University, Case Western Reserve
University, University of Chicago, Drexel University, Fermilab, the
Institute for Advanced Study, the Japan Participation Group, Johns
Hopkins University, the Joint Institute for Nuclear Astrophysics, the
Kavli Institute for Particle Astrophysics and Cosmology, the Korean
Scientist Group, the Chinese Academy of Sciences (LAMOST), Los Alamos
National Laboratory, the Max-Planck-Institute for Astronomy (MPIA),
the Max-Planck-Institute for Astrophysics (MPA), New Mexico State
University, Ohio State University, University of Pittsburgh,
University of Portsmouth, Princeton University, the United States
Naval Observatory, and the University of Washington.



\end{document}